\documentclass[aps,prl,groupedaddress,twocolumn,nolongbibliography]{revtex4-1}
\usepackage{graphicx}
\usepackage{xcolor}

\begin{document}

\title{Telling Solar Neutrinos from Solar Axions When You Can't Shut Off the Sun}

\author{P. Coloma}
\email[]{pilar.coloma@ift.csic.es}
\affiliation{Instituto de F\'isica Te\'orica UAM/CSIC, Calle Nicol\'as Cabrera 13-15, Universidad Aut\'onoma de Madrid, Spain}
\author{P. Huber}
\email[]{pahuber@vt.edu}
\author{J. M. Link}
\email[]{jmlink@vt.edu}
\affiliation{Center for Neutrino Physics, Virginia Tech, Blacksburg, Virginia, USA}

\date{\today}

\begin{abstract}
The XENON1T experiment recently reported an excess of events at low electron recoil energies, which may be due 
to interactions of solar neutrinos inside the detector via a large 
neutrino magnetic moment. We point out that a $^{51}$Cr neutrino 
source placed close to the detector can directly test this 
hypothesis.
\end{abstract}


\maketitle
  
The recent observation of an excess of low-energy electron recoils by
the XENON1T Collaboration~\cite{Aprile:2020tmw} has generated a strong
interest in the particle physics community, since it may be
interpreted as a signal from new physics beyond the standard model
(BSM)\@.  Two of the leading hypotheses, solar axions and solar neutrino
scattering with an anomalously large neutrino magnetic moment
($\nu$MM), would be expected to produce very similar excesses that rise 
rapidly as the electron recoil energy decreases, particularly below 5~keV\@.  
Given the energy smearing and threshold effects at around 2~keV, their 
signatures are essentially indistinguishable in the XENON1T detector.  So 
the questions arises: how do you tell solar neutrinos from solar axions 
when you can't shut off the sun?  The answer comes from the 
GALLEX~\cite{Hampel:1997fc} and SAGE~\cite{Abdurashitov:1998ne} gallium 
detector solar neutrino experiments, which used mega-Curie (MCi) scale 
electron capture sources---mostly $^{51}$Cr~\cite{Cribier:1996cq, 
Hampel:1997fc,Abdurashitov:1998ne,Gavrin:2020} but also 
$^{37}$Ar~\cite{Abdurashitov:2005tb}---as an ``artificial sun'' to test 
their full-system neutrino detection efficiency.  The use of a $^{51}$Cr 
source with XENON1T would definitely show if the excess is due to 
neutrinos, or if some other interpretation, like axions, dark matter or 
tritium decay, is required.  It's worth noting that both the 
axion~\cite{luzio2020solar} and $\nu$MM~\cite{Raffelt:1999gv} interpretations 
are in strong tension with astrophysical constraints.

In 2014, we conducted a study on the physics potential of combining a
liquid Xe (LXe) dark matter detector with a MCi-scale $^{51}$Cr
source~\cite{Coloma:2014hka}, and computed the expected sensitivity to
$\nu$MM\@. A simple scaling from that result shows that a single
exposure of the XENON1T detector to a MCi-scale $^{51}$Cr source
would exceed the neutrino event count observed by XENON1T, in just a 
fraction of the 226.9 days used in their study.  In other words: this 
source run would have a larger signal and just a fraction of the 
background events compared to their data set with the reported excess.  
While the background rate in XENON1T is higher than we assumed in our 
original analysis, our present proposal relies only on improving upon 
the sensitivity of the current XENON1T result in order to disambiguate 
the two BSM hypotheses which are correlated with the solar power.  We 
surmise that the sensitivity to an anomalously large $\nu$MM would 
exceed that required to rule out a $\nu$MM compatible with the excess.
A result compatible with the SM expectation would aid in the 
interpretation of the XENON1T results by rejecting all neutrino
hypotheses.  At the same time it may well be able to improve upon the 
best terrestrial limits on $\nu$MM~\cite{Beda:2013mta}.  On the other 
hand, if the anomaly were to persist in the source data, that would 
strongly support the $\nu$MM hypothesis or any suitable explanation of 
the excess due to novel interactions of solar neutrinos~\cite{boehm2020light, 
bally2020neutrino,sierra2020light,khan2020nonstandard,lindner2020xenon1t}.

\begin{figure}[b]
\includegraphics[width=0.3\textwidth]{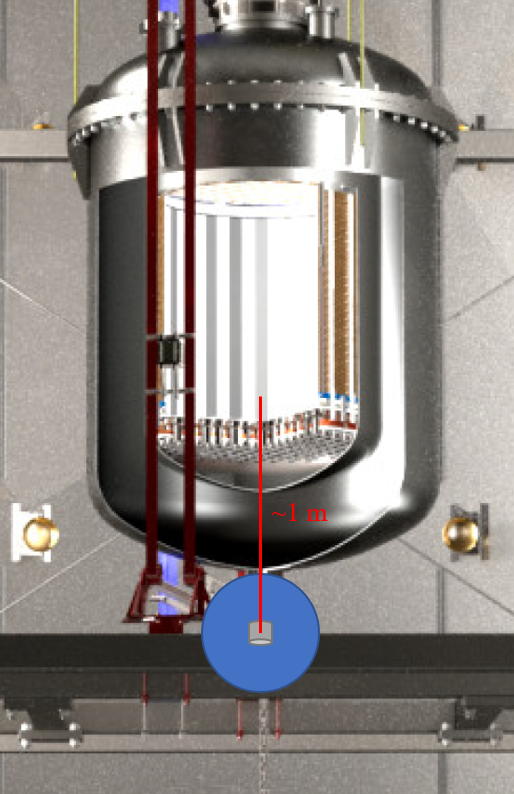}%
\caption{\label{figure1}Drawing of the XENON1T cryostat in its water shield~\cite{Aprile:2017aty} with a possible source bundle---a 10~cm cylindrical source surrounded by a 20~cm thick spherical tungsten shield---superimposed about 1~meter below the bottom edge of the detector fiducial volume.}
\end{figure}

As a specific example, we consider a $^{51}$Cr source with an initial
strength of 3~MCi, which is comparable to, and even slightly smaller
than, the source from Ref.~\cite{Gavrin:2020}.  It is assumed to be
located with its center 1~meter below the bottom edge of the
detector's fiducial volume (see Fig.~\ref{figure1}).
Unlike our original study which assumed a 100 day
run~\cite{Coloma:2014hka}, here we will consider a run of just 50 days,
which, as noted in \cite{Link:2019pbm}, significantly improves the
signal-to-noise ratio.  Based on the 39.96~day lifetime of $^{51}$Cr,
cutting the run time by half, halves the background, while only
reducing the signal by 22\%.  With this configuration, under the
standard model hypothesis, we would expect 82 $\nu-e$ elastic 
scattering events from source neutrinos with recoil energies from 2 to  
30~keV.  During that same period, 9.3 solar neutrino events are expected 
in that energy range (according to the upper panel in Fig.~3 in 
Ref.~\cite{Aprile:2020tmw}). This corresponds to an increase of 
$9.8$~times in the signal-to-noise ratio, and a factor of 2.2 increase 
in the absolute number of $\nu-e$ elastic scattering events compared 
to the 226.9-day XENON1T data set.

\begin{figure}
\includegraphics[width=0.5\textwidth]{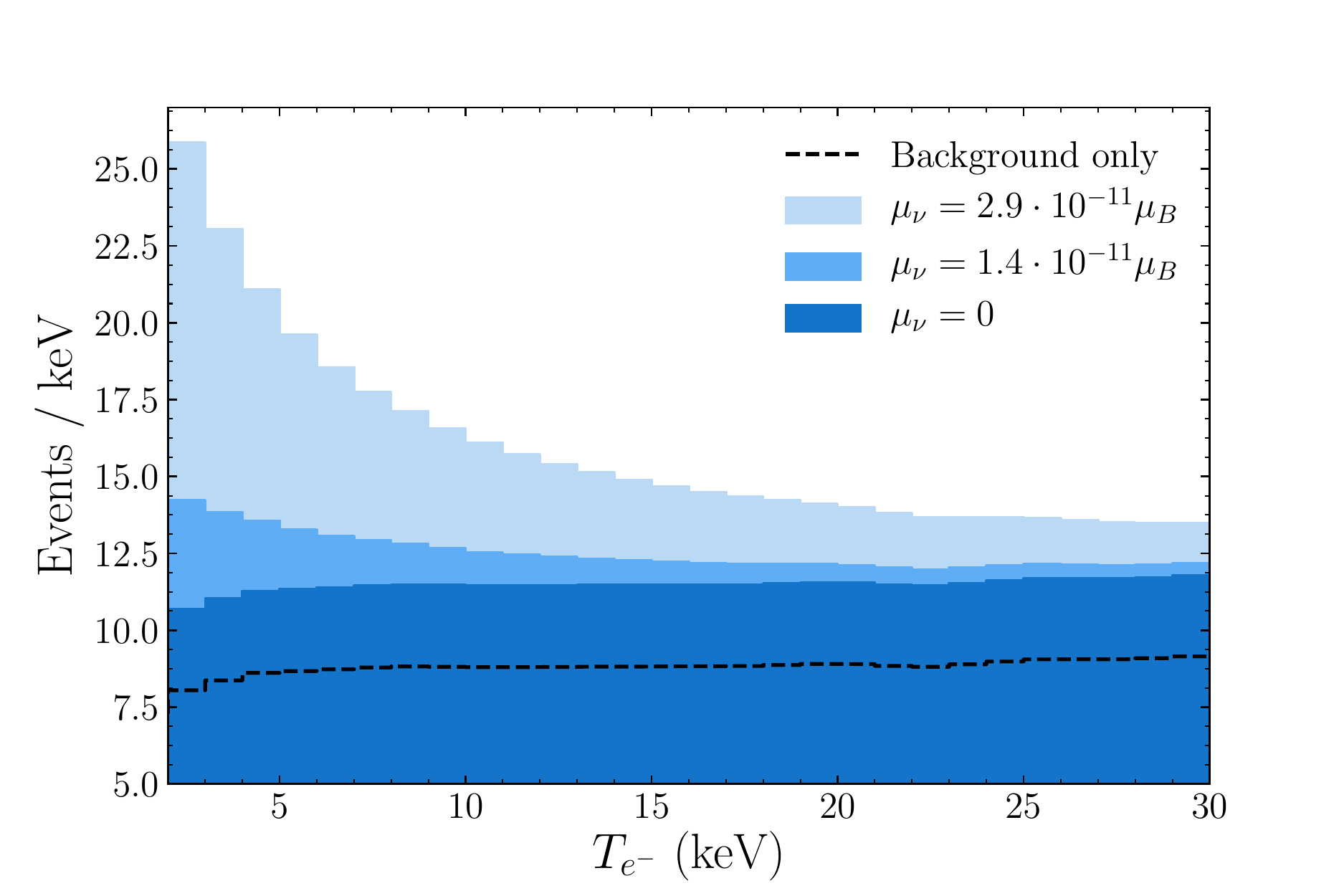}%
\caption{\label{figure2}The expected counts over a 50~day run with an
  initial source strength of 3~MCi.  The background (below the dashed
  line) is scaled from Ref.~\cite{Aprile:2020tmw} and includes solar 
  neutrinos; the dark blue histogram shows the total events with
  source in the standard model ($\mu_\nu=0$); the medium blue shows 
  the additional events with $mu_\nu=1.4\times10^{-11}\,\mu_B$,
  the XENON1T lower limit; and the light blue is the expected signal
  for $\mu_\nu=2.9\times10^{-11}\,\mu_B$ the XENON1T upper limit. }
\end{figure}

Figure~\ref{figure2} shows our projected energy spectrum for three values of  
$\mu_\nu$ which include zero, and the upper and lower bounds 
of the XENON1T 90\% CL allowed range~\cite{Aprile:2020tmw}. We took the energy resolution
to be $44.8\%/\sqrt(E(\rm{keV})$, which was extracted from Fig.~1 in 
Ref.~\cite{Aprile:2020tmw}, a flat efficiency of 90\% above 2~keV and the expected background distribution from Fig.~3 in Ref.~\cite{Aprile:2020tmw}. Across these 28 bins, we find $\Delta\chi^2$s 
relative to $\mu_\nu=0$ of 5.0 and 66.5 for the lower and upper bounds 
respectively. Thus we conclude that the XENON1T preferred range for $\nu$MM would be covered 
at greater than 95\% CL.  

In conclusion, we propose a simple test to distinguish between the two
solar correlated BSM explanations for the XENON1T anomaly by using
existing technology for a MCi-scale $^{51}$Cr source combined with the 
existing XENON1T detector.  A positive detection of an excess would be 
tantamount to the discovery of new physics.  Beyond a test of the $\nu$MM 
hypothesis, combining large LXe dark matter detectors with  MCi-scale 
electron capture sources enables a broad range of BSM physics searches 
including sterile neutrinos~\cite{Coloma:2014hka}, non-standard 
interactions, and light $Z^{\prime}$~\cite{Link:2019pbm}. A detailed 
study of the experimental feasibility and sensitivity would be best 
conducted by the XENON1T collaboration.

\begin{acknowledgments}
  
PC acknowledges support from the Spanish MICINN through the ``Ram\'on
y Cajal'' program with grant RYC2018-024240-I, and from the Spanish
Agencia Estatal de Investigacion through the grant ``IFT Centro de
Excelencia Severo Ochoa SEV-2016-0597''\@.  PH and JML acknowledge 
support from the US Department of Energy under contracts \protect{DE-SC0018327}
and \protect{DE-SC0020262}.
\end{acknowledgments}

\bibliography{refs}
\end{document}